\documentclass[PRL,twocolumn]{revtex4}
\usepackage{amsmath,amssymb,latexsym,epsfig,graphics,epsf}
\usepackage{amsmath}
\DeclareMathOperator{\Tr}{Tr}
\usepackage{fixltx2e} 
\usepackage{graphics}
\usepackage{float}
\usepackage{graphicx}
\usepackage{caption}
\usepackage{multirow}
\usepackage{subfigure}

\newcommand{\be}{\begin{equation}}
\newcommand{\ee}{\end{equation}}
\begin{document}
\title{``IT FROM BIT" \\ How does information shape the structures in the universe?}
\author{S.\ Davatolhagh}
\altaffiliation{Corresponding Author: davatolhagh@shirazu.ac.ir}
\author{A.\ Sheykhi}
\author{M.\ H.\ Zarei}
\affiliation{Department of Physics, School of Science, Shiraz
University, Shiraz 71946, Iran}
\date{\today}
\begin{abstract}
Based on a synthesis of three main ingredients: (i) the Shannon
information in nonequilibrium systems, (ii) the semiclassical
energy-time quantization rule, and (iii) the quasistatic
information-energy correspondence, a new general rule for the
quantization of quasistatic information states supported by an
environment away from equilibrium is introduced if the history of
the environment is known as a function of time in terms of its
thermodynamic potential for information $T(t)\Delta S(t)$ that is
a free energy measuring the distance from equilibrium $\Delta
S(t)$, and $T(t)$ is the mean temperature of the environment at
time $t$. This all new quasistatic information-time quantization
rule is applied to the expanding universe using a phenomenological
thermodynamic potential for information in the matter dominated
era in order to find the eigen-informations of the persistent
structures that are supported by the universe (or the local
environments therein) at any given epoch, thus providing an
information-theoretic foundation for formation of structures and
rise of complexity with time that embodies the cosmic evolution as
epitomized by the late Wheeler's famous conjecture ``{\it it from
bit}". This theoretical procedure must also open new avenues for
further research into the quantum theory of information and
complexity in nonequilibrium thermodynamics.

\end{abstract}
\maketitle

\section{Introduction}\label{sec1}
The most intriguing and fundamental problem in science is probably
the discovery of the very basic laws that govern the emergence and
gradual evolution of structures of increasing complexity since the
inception of time. The term ``cosmic evolution" is often used to
describe the formation of persistent structures of increasing
complexity or information content from the beginning primordial
radiation, called the big-bang, to date including the elementary
particles, nuclei, light atoms, galaxies, stars, planets, and life
itself in the order of higher complexity. The primitive radiation
era gave way to the more complex matter era 380,000 years
following the big bang, and now after 13.7 billion years we are
well within the life era characterized by highly complex life
forms (including us the Homo sapiens) coming to existence and
evolving with the passage of time \cite{Chaisson}.

The above established facts at first may seem to contradict the
second-law of thermodynamics or the law of increasing entropy that
made physicists in the late nineteenth century believe that the
universe must approach a final state of heat-death. The key to
resolving this apparent paradox is to realize that {\em the
universe has never been static but always expanding}. To be more
precise, in any isolated system the total energy content or the
internal energy $U$ is a constant by the first-law of
thermodynamics $\Delta U = Q_{\rm in} + W_{\rm on}$, where $Q_{\rm
in}$ is the net heat intake and $W_{\rm on}$ is the net work done
on the system. Although by the first-law the total energy content
$U$ of an isolated system is constant because no energy in the
form of heat or work can either enter or leave the system, $Q_{\rm
in} = W_{\rm on} = 0$, the second-law of thermodynamics stipulates
that there is a natural tendency for the high-quality energy
available for doing work or the free energy $F$ to run down in an
isolated system such that in the final state of equilibrium free
energy minimizes to a value consistent with the external
constraints: $\Delta F = 0$, and $\Delta^2 F > 0$. Clearly in
equilibrium the free energy is at the rock bottom and all
thermodynamic processes cease to continue. Because free energy
$F=U-TS$ is minimized by maximizing the entropy $S$, the state of
equilibrium is also characterized by the maximum entropy: $\Delta
S =0$, and $\Delta^2 S <0$ \cite{Callen}. Indeed entropy is a
measure of macroscopic uniformity or lack of persistent structures
within the system that may epitomize order. So entropy is a
measure of disorder and reaches its maximum attainable value in
equilibrium consistent with the external constraints imposed on
the system: $S=S_{\rm max}$ in equilibrium. In a static universe,
therefore, it is natural to assume that no matter how far in time,
there comes a time when the universe equilibrates and all natural
processes including life will stop. This was the idea held by the
nineteenth century physicists such as Helmholtz and Clausius who
of course were not aware of the fact that the universe is not
static but expanding. The discovery of cosmic expansion in 1920s
by Hubble and Lema\^{i}tre among others, had far reaching
implications not only for modern cosmology but also for negating
the cosmic heat-death scenario {\em if the expansion rate of the
universe exceeds the rates of reactions and processes responsible
for equilibration}. The cosmic expansion is succinctly summarized
in Hubble's law $v=Hd$, which states that the recession speed of a
galaxy $v$ is proportional to the distance from observer $d$. The
proportionality parameter $H$ called the Hubble parameter is of
the order of the inverse of the age of the universe $H\sim
t^{-1}$.

The nonequilibrium nature of the expanding universe is not only
responsible for the formation of persistent living organisms in
the later stages of cosmic evolution but also for the birth of the
tiniest of structures as fundamental as elementary particles in
the early universe. Considering the birth of the elementary
particles, nucleons, light nuclei, and light atoms in the early
universe as entities independent of the primordial radiation, it
is the generally accepted view among astroparticle physicists that
\cite{Pathria}
\\ ``{\it As the universe expanded and cooled, the cooling rate
was proportional to the Hubble parameter, that is, of the order of
the inverse of the age of the universe at that point in expansion.
This led to a sequence of important events when different
particles and interactions fell out of equilibrium with the gas of
blackbody photons {\rm [primordial radiation]}. The neutrino and
neutron-proton conversion reactions {\rm [weak interaction]} fell
out of equilibrium at $t\approx 1$ second. Nuclear reactions {\rm
[strong interaction]} that formed light nuclei fell out of
equilibrium at $t\approx 3$ minutes. Neutral atoms fell out of
equilibrium at $t\approx 380,000$ years. All these degrees of
freedom froze out when the reaction rates that had kept them in
equilibrium with the blackbody photons fell far below the cooling
rate of the expanding universe.}" Clearly, a necessary driving
force behind cosmic evolution is the universal expansion that not
only averts the heat-death of a static universe but also creates
the nonequilibrium conditions required for formation of structures
and rise of complexity within the universe and the local
environments therein \cite{Chaisson, Layzer}.

In fact every nonequilibrium environment has an inherent {\em
thermodynamic information content}, which according to Shannon's
definition of information is rigorously quantified by the distance
of the environment from the state of thermodynamic equilibrium or
maximum entropy \cite{Layzer}: \be \Delta S(t) \equiv S_{\rm
max}(t) - S(t)\,.\ee $S(t)$ is the actual entropy of the
environment be it the universe or a local environment therein, and
$S_{\rm max}(t)$ is the maximum attainable or equilibrium entropy
of the environment consistent with the external constraints at the
same time $t$. The details of this quantification of information
content in terms of the deviation from the state of thermodynamic
equilibrium, is presented in the next section following the line
of argument given in Ref.\ \cite{Chaisson}. It must be emphasized
that Eq.\ (1) is in essence the Shannon's measure of the
information content of a message received about the thermodynamic
system, $\Delta I = \log_2\Omega_{\rm i} - \log_2\Omega_{\rm f}$,
that has the effect of reducing the number of possible microscopic
configurations of the system from $\Omega_{\rm i}$ to $\Omega_{\rm
f}$ \cite{Shannon}. Here the logarithm of base two is chosen to
give the information content in units of binary bits. In Eq.\ (1)
{\em the equilibrium state forms the vacuum of information} as it
cannot support any persistent structures but transient equilibrium
fluctuations. Eq.\ (1) also embodies the Layzer's definition of
{\em order as the lack of disorder} such that every time a
thermodynamic system deviates from the state of maximum entropy
(disorder), it must inherently hold a quantifiable amount of
Shannon information (order) \cite{Layzer}. Hence, {\em although in
general the entropic measure of disorder $S(t)$ is rising by the
second-law of thermodynamics, the informational measure of order
$\Delta S(t)$ can also increase if $S_{\rm max}(t)$ is rising at a
rate faster than $S(t)$} (see, e.g., Fig.\ 1). This is believed to
be the basis for increasing order, information content, or
complexity that so vividly characterizes our expanding universe
\cite{Chaisson, Layzer}.

Therefore, there are already clear signs backed by empirical data
as well as theoretical physics that the information content (bit)
inherent to the nonequilibrium universe or local environments
therein such as ours, must play a fundamental role in the
formation and evolution of persistent structures (it), or to use
the catch phrase prevalent among physicists due to Wheeler ``{\em
it from bit}" \cite{Wheeler}. Now a question that immediately
arises is how to develop a quantitative theory based on
information that employs this presumably fundamental quantity
presented by Shannon information, $\Delta S(t)$, to account for
the cosmic evolution at a quantitative level and calculate the
persistent information states supported by the nonequilibrium
universe at different epochs that are so vividly reflected in both
animate and inanimate physical species with {\em discrete} levels
of complexity. In a serious attempt to this end we present a
phenomenological top-down approach to the problem of cosmic
evolution based on information that is of course independent of
the microscopic bottom-up views such as the inflationary theories
\cite{Guth}, and the more recent rival theories based on cyclic
cosmology \cite{Stein}, in that here we treat information as the
most fundamental quantity by employing (i) the thermodynamic
information content of Eq.\ (1) in the appropriate form of an
information free energy $T(t)\Delta S(t)$ (see, also, Fig.\ 1)
where $T(t)$ is the mean temperature of the environment at time
$t$, (ii) the energy-time quantization rule rigorously derived
from Wentzel-Kramers-Brillouin (WKB) semiclassical approximation
of quantum mechanics for bound systems, and (iii) the quasistatic
information-energy correspondence that is inferred from Landauer's
quasistatic limit \cite{Landauer}, together with Szil\'{a}rd's
model of a quasistatic information heat engine \cite{Szilard},
both of which have found experimental validity in the recent years
\cite{exp1,Berut,exp12,exp2,exp3}.

The rest of this paper is organized as follows. In Sec.\
\ref{sec2} the semi-empirical theory leading to the quasistatic
information-time quantization rule is developed based on the above
three ingredients. The method is then applied to the expanding
universe in Sec.\ \ref{sec3} and the resulting quasistatic
information states are compared with the general phenomenology of
the cosmic evolution at a quantitative level. The paper is
concluded with a summary, concluding remarks and directions for
further research in Sec.\ \ref{sec4}.

\section{Quasistatic information-time quantization rule}\label{sec2}
In this section we expand on the three main ingredients referred
to in the introduction in order to derive the quasistatic
information-time quantization rule from their synthesis.

\subsection{Shannon information in nonequilibrium systems}
A fundamental quantity in the definition of Shannon information is
the surprisal $s$, which quantifies the degree of surprise or
amazement in observing a random event. The surprisal of an event
with probability $p$ is proportional to the logarithm of the
inverse probability \be s=K\log\frac{1}{p}=K\log\Omega\ee such
that the surprisal associated with a probable event ($p\rightarrow
1$) is small while that of an improbable event ($p\rightarrow 0$)
is large. The logarithmic nature of surprisal is due to its
additive property when applied to the compound probabilities of
multiple independent events \cite{Shannon}. $K$ is a
proportionality constant that determines the units or
alternatively the units can be set by choosing a particular base
for the logarithmic function. For equally likely events the
probability of any one event is $p=1/\Omega$, where $\Omega$ is
the number of all possible outcomes. It is worth noting that the
surprisal for equally likely events is already beginning to look
like the Boltzmann entropy for the microcanonical statistical
ensemble $S=k_{\rm B}\ln\Omega$. Now according to Shannon the
information content of a message received that reduces the total
number of possible initial outcomes $\Omega_{\rm i}$ to a final
number $\Omega_{\rm f}$ ($< \Omega_{\rm i}$), is given by the
difference between initial and final surprisal: \be \Delta I =
K\log\Omega_{\rm i} - K\log\Omega_{\rm f} = K\log\frac{\Omega_{\rm
i}}{\Omega_{\rm f}}\,. \ee For example in tossing a fair dice the
initial surprisal is $s_{\rm i}=K\log 6$ (in arbitrary units)
because all six outcomes are equally likely. Now if we receive a
message that reduces the number of possible outcomes to four by
informing us that the outcome of the trial is a number less than
five, the final surprisal reduces to $s_{\rm f} = K\log 4$ (in
arbitrary units). Therefore in this example the information
content of the message received according to Shannon is $\Delta I
= K\log 6 - K \log 4 = K\log 1.5$. Equivalently if the initial
probability distribution $\{p_{\rm i}\}$ is reduced to a final
distribution $\{p_{\rm f}\}$ as a result of the message received,
the information content of the message is measured in terms of the
difference between the {\em average surprisal} of initial and
final distribution \cite{Shannon}:
\begin{eqnarray}\label{Shannon} \Delta I &
= & -K\langle\log p_{\rm i}\rangle -(-K\langle\log p_{\rm f}\rangle) \nonumber \\
         & = & -K\sum p_{\rm i}\log p_{\rm i} - (-K\sum p_{\rm f} \log p_{\rm
f})\,.
\end{eqnarray}
The above definition of Shannon information manifestly is the
Gibbs entropy of the initial probability distribution minus that
of the final reduced distribution as a result of the message
received. {\em This reduction in entropy or missing information
quantifies the amount of Shannon information made available.}

Similarly one can measure the information content of any
thermodynamic system in terms of the difference between two
entropies: one of the maximum attainable entropy consistent with
the external constraints or the entropy of equilibrium state
characterized by the probability distribution of microstates
$\{p_{\rm max}\}$, and the other of the actual entropy of the
system with a distribution $\{p\}$:
\begin{eqnarray*}
  \Delta S &=& -k_{\rm B}\sum p_{\rm max} \ln p_{\rm max} - (-k_{\rm B} \sum p \ln p) \\
   &=& S_{\rm max} -S\,,
\end{eqnarray*}
where $k_{\rm B}$ is the Boltzmann constant. This is indeed Eq.\
(1) first proposed by David Layzer \cite{Layzer}, and elaborated
further by Eric Chaisson \cite{Chaisson}. Clearly, {\em
information has the effect of reducing the entropy from its
maximum attainable or equilibrium value}. As a way of
demonstration we imagine entering a messy and disordered room that
is high in entropy or missing information. It is difficult to know
where things are therefore low in information content. On the
other hand a tidy and ordered room is high in information content
because we know where things are--cloths are in the closet, books
are on the shelves, etc., thanks to the sentient agent that tidied
up the room at the first place. So this definition of information
content in terms of distance from equilibrium is indeed consistent
with our intuitive notions of order and information available.
Furthermore, as pointed out in the introduction, Eq.\ (1)
identifies the equilibrium state with the vacuum of information
$\Delta S =0$. Indeed the state of thermodynamic equilibrium is
the simplest state imaginable because it is macroscopically
uniform and devoid of any structures, which means that it can be
simply described by a handful of thermodynamic degrees of freedom
such as a homogeneous temperature, a homogeneous pressure, and
homogeneous chemical potentials of constituent particles. But the
description of a nonequilibrium system tends to be more complex
for it involves the spatial gradients and higher-order derivatives
of the thermodynamic variables as well as their time evolutions.
So the equilibrium state as the vacuum of information, $\Delta S
=0$, is consistent with our intuitive notion of {\em simplicity}
while nonequilibrium states with their inherent information
contents, $\Delta S >0$, tend to exhibit more {\em complexity}.

\subsection{Semiclassical energy-time quantization rule}
In quantum mechanics the incompatible observables whose products
have the physical dimensions of {\em action} and satisfy
commutation relations of the form $[\hat{p},\hat{q}]=-i\hbar$,
where $\hat{p}$ is the momentum operator conjugate with the
coordinate $\hat{q}$, not only satisfy uncertainty relations of
the form $\Delta p \Delta q \geq \hbar /2$, but also constituted
heuristic quantization relations of the form $\oint p dq = nh$ for
bound systems in the spirit of Bohr-Wilson-Sommerfeld old quantum
theory, until a mathematically rigorous derivation of such
quantization relations were obtained from Schrodinger wave
equation through the WKB semiclassical approximation
\cite{Sakurai}: \be\label{WKB} \int_{x_1}^{x_2} p(x) dx =
(n+\delta) \pi \hbar\,. \ee In Eq.\ ({\ref{WKB})
$p(x)\equiv\sqrt{2m(E-V(x))}$ is the magnitude of momentum, $V(x)$
is the potential energy function, $x_1$ and $x_2$ are the
classical turning points of the bound system such that
$E=V(x_1)=V(x_2)$, $n=1,2,\cdots$ is a positive integer, $\delta$
is a system-dependent fractional constant in the range $-1<\delta
< 1$, and $\hbar = h/2\pi$ with $h$ the Planck's constant.
Therefore, it is natural to conclude that there must be a
semiclassical energy-time quantization rule consistent with the
WKB approximation Eq.\ (\ref{WKB}), which is of considerable
interest to our analysis. In fact by simply applying a change of
variables from position to time in Eq.\ (\ref{WKB}) \be
\int_{x_1}^{x_2} p(x) dx = \int_{t_1}^{t_2} p(x(t)) \dot{x}
dt\,,\ee and using $p\equiv \sqrt{2m(E-V(x(t))}=m\dot{x}$ for the
momentum together with $E-V(x(t))=m\dot{x}^2/2$ for the
instantaneous kinetic energy, we readily find the right
semiclassical energy-time quantization rule: \be\label{ET}
\int_{t_1}^{t_2}\left[E_n - V(x(t))\right] dt =
\frac{1}{2}(n+\delta)\pi\hbar.\ \ee In the above semiclassical
energy-time quantization rule, Eq.\ (\ref{ET}), $E_n$'s are the
semiclassical energy eigen-values, and $V(x(t))$ is the potential
energy function expressed as a function of time. Of all possible
trajectories, $x(t)$ represents the classical or least-action
trajectory of the mechanical system in question, and the time
limits $t_1$ and $t_2$ are the classical turning times when $E_n =
V(x(t_1))=V(x(t_2))$. Indeed the application of Eq.\ (\ref{ET})
together with the virial theorem to the system of harmonic
oscillator and particle in a box with $\delta = -1/2$ and $\delta
= 0$, respectively, results in the exact energy eigen-values being
recovered. Eq.\ (\ref{ET}) also is validated by testing it for
other potential energy wells to obtain the semiclassical energy
eigen-values that are born out of Eq.\ (\ref{WKB}). It is worth
noting that recently a similar semiclassical energy-time
quantization rule of the form $\oint E dt = 2\pi \hbar (n+\beta)$
has been derived for extreme relativistic particles  characterized
by linear energy-momentum relation $E=pc$, where $c$ is the speed
of light (in relativistic units $c=1$) \cite{Dolce}. We note that
exactly the same result follows from WKB semiclassical
approximation Eq.\ (\ref{WKB}) with our procedure of changing
variables from position to time $dx = \dot{x} dt$ and further
noting that $\dot{x} =c =1$ (in relativistic units). Finally it
must be pointed out that Eq.\ (\ref{ET}) is subject to the same
limitations on the WKB semiclassical approximation Eq.\
(\ref{WKB}) from which it derives. However when applied
judiciously to bounded systems they can greatly simplify the
eigen-energy problem albeit at the cost of some accuracy. The WKB
approximation has also been used to handle scattering problems
\cite{Griffiths}.

As a way of demonstrating the utility of semiclassical energy-time
quantization rule, Eq.\ (\ref{ET}), here we consider the canonical
example of the harmonic oscillator potential energy well $V(x) =
\frac{1}{2}m\omega^2x^2$. Considering the classical or
least-action trajectory $x(t)=x_0\cos(\omega t)$, the potential
energy as a function of time becomes $V(x(t)) =
\frac{1}{2}m\omega^2x_0^2\cos^2(\omega t)$, which on substitution
in Eq.\ (\ref{ET}) gives \begin{eqnarray}\label{SHO}
  \frac{1}{2}(n+\delta)\pi\hbar &=& \int_0^{\frac{T}{2}}\left[E_n - \frac{1}{2}m\omega^2x_0^2\cos^2(\omega t)\right]dt \nonumber \\
   &=& E_n\,\frac{T}{2} - \frac{1}{2}m\omega^2
x_0^2\,\frac{T}{4}\,.
\end{eqnarray}
On solving Eq.\ (\ref{SHO}) for quantized energies $E_n$, and
noting that the time period $T=2\pi/\omega$, we get
\begin{eqnarray}\label{SHO2}
  E_n &=& (n+\delta)\frac{\hbar\omega}{2} + \frac{1}{4}m\omega^2x_0^2 \nonumber\\
   &=& \langle{\rm K.E.}\rangle + \langle{\rm P.E.}\rangle\,.
\end{eqnarray}
In Eq.\ (\ref{SHO2}) the second term on the right hand side is
manifestly the average potential energy, and therefore the first
term must be the average kinetic energy. By virial theorem
therefore $\langle{\rm P.E.}\rangle = \langle{\rm K.E.}\rangle
=(n+\delta)\hbar\omega/2$, which immediately gives the exact
eigen-energies $E_n=(n+\delta)\hbar\omega$ for harmonic oscillator
with $n=1,2,\cdots$ and $\delta=-1/2$.

\subsection{Quasistatic information-energy correspondence} At this
point we invoke the quasistatic information-energy correspondence
that is based on empirically proven facts embodied in both
Landauer's quasistatic limit that specifies the minimum amount of
{\em heat} generated by erasing a binary bit of information in the
limit of long erasure cycles, \be\label{Qbit} Q({\rm
1\,bit})=k_{\rm B}T\ln 2\ee ($T$ being the ambient temperature)
\cite{Landauer}, and a well-studied model of a quasistatic
information heat engine due to Szil\'{a}rd that implies a bit of
information can be converted to {\em work} or free energy
equivalent of the Landauer's limit without contradicting the
second law of thermodynamics \cite{Szilard}: \be\label{Wbit}
W({\rm 1\,bit})=k_{\rm B}T\ln 2\,.\ee Indeed more recent
experiments conducted on double-well potential systems as binary
bits of information have unequivocally proven the validity of Eq.\
(\ref{Qbit}) for quasistatic erasure cycles
\cite{exp1,Berut,exp12}. Similarly recent state-of-the-art
observations of microscopic particles in stairwell potentials have
given definitive proof of Eq.\ (\ref{Wbit}) \cite{exp2, exp3}.

Although in microscopic systems the quantities such as heat, work,
energy, etc., fluctuate and stochastic violations of the
second-law have been recorded \cite{exp4}, but on average the
second-law of thermodynamics holds well such that according to the
Jarzynski equality \cite{Jarzynski} \be\label{2nd}\langle\Delta F
-W \rangle\le0\,,\ee where $F$ and $W$ are the stochastic free
energy and mechanical work, respectively, and the angular brackets
denote the ensemble average such that all thermodynamic quantities
are obtained from ensemble averaging. According to this statement
of the second-law of thermodynamics, Eq.\ (\ref{2nd}), the
mechanical work done on the system by an external force tends to
increase the free energy of the system. The less than sign means
that some of the energy transferred to the system in the form of
work is lost by entropy production. The equality therefore only
applies to quasistatic processes where all of the external work
adds up to the free energy of the system.

It is a matter of considerable interest that the information
obtained about a microscopic system through measurements can also
be used as a resource for free energy such that the system can
gain free energy larger than the thermodynamic mechanical work
$W_{\rm on} \equiv\langle W\rangle$ applied to it. Thus the
generalized second-law that includes the effect of information
must read \cite{exp5} \be\label{2ndg}\langle\Delta F \rangle -
W_{\rm on} \le k_{\rm B}TI\ln 2\ee  where, $I$ represents the bits
of information obtained by the measurements carried out on the
system via a feedback control mechanism, and the equal sign only
holds for the quasistatic information heat engines. The
generalized second-law, Eq.\ (\ref{2ndg}), asserts that in
addition to mechanical work, the information obtained by observing
the system can be used to increase its free energy through a
feedback control mechanism. As an example of how information can
be used to raise the free energy in the absence of any mechanical
work, consider a microscopic particle in a potential energy
staircase at ambient temperature. Due to the finite temperature
the particle sometimes jumps up the steps as a result of the
random thermal fluctuations although the general tendency is to
run down the staircase in order to minimize its free energy
consistent with the external constraints. Now if there is a
sentient being (such as a fairy, if not a demon!) that blocks the
downward path of the particle every time it observes an upward
jump, the free energy of the particle increases just through the
information obtained in a feedback control procedure, without any
mechanical work actually done on the system.

Now because a {\em bit} of information can be quasistatically
converted to a well-defined amount of heat $Q$ and work $W$ given
by Eq.\ (\ref{Qbit}) and Eq.\ (\ref{Wbit}), respectively, it must
find correspondence with the {\em energy} $E$ in the limit of
quasistatically slow processes through the first-law of
thermodynamics: \be\label{1st}\langle\Delta E\rangle=Q_{\rm
in}+W_{\rm on}\ee where, $\langle\Delta E\rangle$ is the change in
internal energy, and $Q_{\rm in}\equiv\langle Q\rangle$ the
thermodynamic heat injected into the system. Hence, {\em a bit of
information can raise the energy $E$ by a well-specified amount
through either quasistatic heat or work}. It must be emphasized
that internal energy $U=\langle E_n\rangle$ is nothing but the
ensemble average of the eigen-energies, which are obtained at a
semiclassical level from the energy-time quantization rule Eq.\
(\ref{ET}). It is worth noting that a similar correspondence
between digital information and energy has been proposed in Ref.\
\cite{Melvin} under the title of mass-energy-information
equivalence principle.

This empirical {\em quasistatic information-energy correspondence}
therefore justifies the generalization of semiclassical
energy-time quantization rule, Eq.\ (\ref{ET}), to an all new {\em
quasistatic information-time quantization rule}\be\label{IT}
\int_{t_n}^{t_0}\left[I_n - V(t)\right] dt =
\frac{1}{2}(n+\delta)\pi\hbar\,, \ee where $|I_n|$'s are the
information eigen-values in units of energy, and $V(t)$ is given
by the thermodynamic potential for information $T(t)\Delta S(t)$,
which sets an upper bound (or maximum) for the information content
at any time $t$ during the history of the environment:
\be\label{IP} V(t)= -T(t)\Delta S(t)\,. \ee It must be noted that
$T(t)\Delta S(t)$ is a genuine thermodynamic potential for
information because not only it has the dimensions of free energy
but also on minimization it gives the state of thermodynamic
equilibrium characterized by $\Delta S = S_{\rm max}-S=0$
\cite{Callen}. The above Eq.\ (\ref{IT}), together with Eq.\
(\ref{IP}), constitute an all new quasistatic information-time
quantization rule generally applicable to any slowly varying
nonequilibrium thermodynamic system whose thermodynamic potential
for information can be determined accurately as a function of time
in order to obtain the eigen-informations of the quasistatically
evolving physical objects supported by the medium. Hence a
significant advantage of Eq.\ (\ref{ET}) over Eq.\ (\ref{WKB}) is
that it can be adapted for application to situations beyond simple
mechanical systems by employing a time-dependent thermodynamic
potential for information in nonequilibrium environments, Eq.\
(\ref{IP}), with the right free energy dimensions, and the
interpretation of eigen-informations $|I_n|$'s as the quantized
information contents of the persistent physical structures that
are supported by the out-of-equilibrium environment.

With application to the expanding universe in mind, in Eq.\
(\ref{IT}) $t_0$ is the present epoch or the age of the universe,
and $t_n$ is given by $I_n = V(t_n)$ that is comparable with the
time that the information state $I_n$ first appears. By picking a
different epoch in place of $t_0$, Eq.\ (\ref{IT}) can be applied
to different stages of the cosmic evolution including the future
ones if the thermodynamic potential for information in Eq.\
(\ref{IP}) can be predicted for the foreseeable future. It must be
noted that $|I_n|$'s have the dimensions of (informational) energy
and that the information eigen-values in units of bits are given
by $|I_n|/k_{\rm B}\bar{T}\ln 2$, where $k_{\rm B}\bar{T}\ln 2$ is
the quasistatic energy equivalent of one bit of information (the
Landauer's limit), and $\bar{T}$ is the average temperature of the
environment that can be estimated from $\bar{T} \simeq
\int_{t_n}^{t_0} T(t)dt/(t_0 - t_n)$. $t_0 - t_n $ is a good upper
bound for the age of the information state. It must be emphasized
that Eq.\ (\ref{IP}) sets the {\em maximum informational energy}
that a system may have at any given time through the thermodynamic
potential for information $T(t)\Delta S(t)$, and plays the role of
a bound for quantization in Eq.\ (\ref{IT}) similar to the {\em
minimum mechanical energy} set by the potential energy function
$V(x(t))$ in Eq.\ (\ref{ET}).

It must also be pointed out that the semiclassical
information-time quantization rule Eq.\ (\ref{IT}), must be
regarded as a first step towards a quantum mechanical theory of
information in nonequilibrium thermodynamics as it predicts
eigen-informations belonging to an as yet to be determined
information operator, which may well be linked to the density
operator in the form of $-\ln\rho$ whose ensemble average gives
the von Neumann entropy: $S=-\Tr \rho\ln\rho=-\sum_i \lambda_i
\ln\lambda_i$. Furthermore, a quantum phase processing algorithm
has been recently introduced that also extracts the
eigen-information of the quantum systems by measuring the ancilla
qubit \cite{WZYW}. Finally, we must also point out that another
definition of information states has been recently introduced in
Ref.\ \cite{Vopson2023} that although instructive and very useful
in its own right, does not directly relate to the information
eigen-values introduced here.

\begin{figure}[H]
\centering
\includegraphics[width =0.45\textwidth]{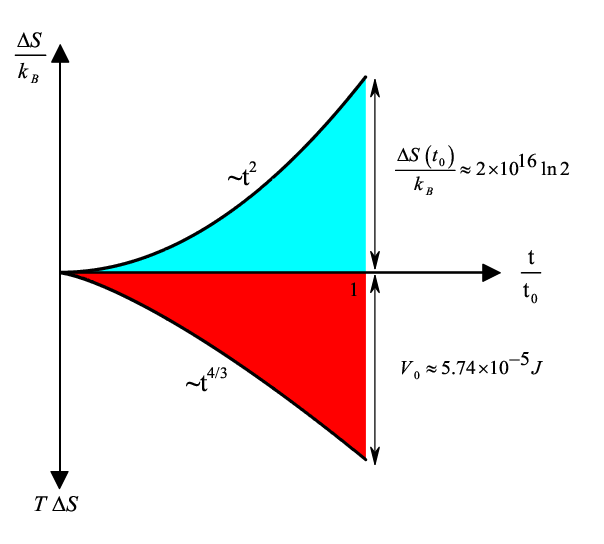}
\caption{The deviation of the universe (or a local environment
therein such as ours) from the state of equilibrium $\Delta
S(t)/k_{\rm B}= 2\ln 2\times 10^{16} (t/t_0)^2$ is shown for the
matter dominated era with the parameters determined
semi-empirically, and highlighted in blue. The growth of the
thermodynamic potential for information $T(t)\Delta
S(t)=5.74\times 10^{-5}(t/t_0)^{4/3}$\,J also is shown and
highlighted in red that acts as a bound for the quantization of
quasistatic information states (see text for details).}
\label{Fig1}
\end{figure}

\section{Application to expanding universe}\label{sec3}
In order to apply the quasistatic information-time quantization
rule, Eq.\ (\ref{IT}), one needs to know the history of the
environment in terms of the thermodynamic potential for
information in Eq.\ (\ref{IP}). So we begin by deriving a
phenomenological expression for the thermodynamic potential for
information in matter dominated era by employing the holographic
principle and taking note of the fact that the most complex
phenomenon of the present epoch is the human brain.

The phenomenologically derived expression for the deviation from
equilibrium $\Delta S(t)$ (which became significant following the
end of the radiation era \cite{Chaisson}), is pictorially depicted
in Fig.\ 1 for the matter dominated era that began 380,000 years
following the big-bang: \be\label{dst} \frac{\Delta S(t)}{k_{\rm
B}}= 2\ln 2\times 10^{16} \left(\frac{t}{t_0}\right)^2.\ee To
arrive at this expression, the holographic principle of the
quantum gravity \cite{Hologram}, and the upper bound for the
equilibrium entropy imprinted on the boundary of the observable
universe given by Bekenstein-Hawking formula \cite{Bekenstein,
Hawking}, are used to determine the time-variation of the maximum
attainable entropy \be\label{smax} \frac{S_{\rm max}}{k_{\rm B}}=
\frac{A}{4 l_{\rm P}^2}\sim t^2\,.\ee In Eq.\ (\ref{smax}),
$A=4\pi\tilde{r}^2$ is the area of the apparent horizon with
$\tilde{r} = c/H$ the cosmic radius of the observable flat
universe, and $l_{\rm P}\equiv\sqrt{\hbar G/c^3}$ is the Planck's
length. The Hubble parameter is given by
$H\equiv\dot{a}/a=\frac{2}{3t}$ in the matter dominated era
because the cosmic scale factor varies with the time as $a(t)\sim
t^{\frac{2}{3}}$. Hence, the maximum attainable or equilibrium
entropy of the universe is expected to vary with the second power
of time in the matter dominated era as given by Eq.\ (\ref{smax}).
The actual entropy $S(t)$, and hence $\Delta S(t)$ shown in Fig.\
1, also are taken to have a similar dependence on time but with
lower proportionality constants that are chosen phenomenologically
by requiring that at the present epoch $\Delta S (t_0)/k_{\rm
B}\approx 2\times 10^{16}\ln 2$, which corresponds to the
information content of the most complex phenomenon of our present
time, namely, the human brain (more of which later) \cite{brain1}.
Furthermore, the cosmic temperature in matter dominated era varies
like $T(t)\sim a^{-1}(t)\sim t^{-\frac{2}{3}}$. Hence the
time-variation of the thermodynamic potential for information in
the matter dominated era becomes \be\label{4/3} T(t)\Delta S(t)
\sim t^{\frac{4}{3}}\ee with a power-law exponent of four-thirds
that is also depicted in Fig.\ 1. It must be noted that
determining the precise expression for the actual entropy $S(t)$
from first principles, and therefore that of the distance from
equilibrium $\Delta S(t)$, is almost impossible because the exact
nature of much of the matter and the energy in the universe in the
form of dark matter and dark energy is unknown. So one has to
resort to reasonable assumptions and phenomenological numerical
parameters to determine the thermodynamic potential for
information as accurately as possible.

In view of the power-law nature of the thermodynamic potential for
information in matter dominated era, Eq.\ (\ref{4/3}), we begin by
considering a generic information well of the power-law form
\be\label{alpha} V(t) = -V_0 \left(\frac{t}{t_0}\right)^{\alpha}\
\ \ \ \ (\alpha> 0).\ee $V_0$ is a phenomenological parameter
denoting the information well-depth at the present epoch the value
of which must be determined by requiring that on substitution in
Eq.\ (\ref{IT}) the highest information eigen-value $|I_1|$
matches the most complex phenomenon of the present epoch, which is
almost always taken to be the human brain. The complexity of a
species, rather similar to its information content \cite{entropy},
is usually attributed to the number of accessible states in its
environmental response or cognitive bandwidth \cite{Adam}. For
humans this is often quantified in terms of the brain information
content, which in more recent estimates is the equivalent of 2.5
million gigabytes or $2\times 10^{16}$ bits of digital memory
\cite{brain1}, although  other estimates  differ by few orders of
magnitude ranging from $10^{14}$--$10^{18}$ bits. For practical
purposes, therefore, the information well-depth at the present
epoch is given by \be\label{v0} V_0 \approx 2\times 10^{16} k_{\rm
B}T\ln 2 = 5.74\times 10^{-5}\,{\rm J}\,,\ee where $T=300$\,K is
the ambient temperature of the environment within which the brain
operates. This is the information well-depth at the present epoch,
also shown in Fig.\ 1.

\subsection{Early stages}
Before application to the present epoch, Eq.\ (\ref{IT}) is solved
for the information eigen-values in a range of times that the
universe has just begun to deviate from equilibrium in order to
demonstrate the sequential birth of the quasistatic information
states and their subsequent dynamical evolution. To obtain the
eigen-informations for an arbitrary epoch $\tau$, the upper limit
$t_0$ in Eq.\ (\ref{IT}) must be replaced by $\tau$ and the
integral performed with $V(t)$ given by Eq.\ (\ref{alpha}). The
lower limit $t_n$ satisfies $I_n = V(t_n)$, which solves to give
$t_n = t_0 (|I_n|/V_0)^{1/\alpha}$. On carrying out the integral,
the following algebraic equation for the information eigen-values
$x_n\equiv -I_n = |I_n|$ is obtained: \be\label{algebra}
\frac{\alpha}{\alpha
+1}\left(\frac{x_n}{V_0}\right)^{\frac{\alpha+1}{\alpha}}-\frac{\tau}{t_0}
\left(\frac{x_n}{V_0}\right) + \frac{1}{\alpha +1}
\left(\frac{\tau}{t_0}\right)^{\alpha + 1} - \frac{\tilde{n}}{t_0
V_0}=0\,\ee where $\tilde{n}\equiv\frac{1}{2}(n+\delta)\pi\hbar$.
In general Eq.\ (\ref{algebra}) must be solved numerically for the
eigen-informations $x_n(\tau)$ at different epochs. But it solves
exactly for the case $\alpha=1$ or the linear information well,
which we consider for the sake of illustrating the {\em dynamics}
of quasistatic information states, as it reduces to a second-order
algebraic equation with the solutions \be\label{xnt} x_n(\tau) =
V_0\tau/t_0\pm\sqrt{V_0(n+\delta)\pi\hbar /t_0}\,.\ee In Eq.\
(\ref{xnt}) the negative sign gives the physical solutions because
the allowed eigen-informations cannot exceed the information
well-depth at the epoch $\tau$ being considered:
$0<x_n(\tau)<|V(\tau)|$. It must be noted that the first term in
Eq.\ (\ref{xnt}) is nothing but the information well-depth,
$|V(\tau)|=V_0 (\tau/t_0)^{\alpha}$, with $\alpha=1$ for the
linear information well. The maximum number of eigen-informations
supported by the universe as a function of epoch, $n_{\rm
max}(\tau)$, is obtained by letting $x_n=0$ in Eq.\
(\ref{algebra}) for the generic power-law information well with
the result \be\label{nmax} n_{\rm
max}(\tau)=\frac{2\,|V(\tau)|\,\tau}{\pi\hbar(1+\alpha)} - \delta
\ee (or the integer part thereof). Clearly, in the case of linear
information well, $\alpha=1$, the number of quasistatic
information states increases with the second power of $\tau$,
while their information contents or complexities increase linearly
with $\tau$. When does the $n^{\rm th}$ state first appear in the
age of the universe? $\tau_n$ is obtained by substituting $n$ for
$n_{\rm max}$ and $\tau_n$ for $\tau$ in Eq.\ (\ref{nmax}), which
solves to give \be\label{taun} \tau_n = t_0\left[\frac{\pi\hbar
(1+\alpha)}{2t_0V_0}\,(n+\delta)\right]^{\frac{1}{1+\alpha}}. \ee
For $\alpha=1$ or the linear information well, Eq.\ (\ref{taun})
becomes $\tau_n = \sqrt{(n+\delta)\pi\hbar t_0/V_0}\approx
\sqrt{n}$\,($1.58\,\mu$s), where the numerical substitutions
$\hbar=1.055\times10^{-34}$\,J.s, $t_0=
13.7\times10^9$\,yrs\,$=4.32\times10^{17}$\,s, $V_0\approx
5.74\times 10^{-5}$\,J, and $\delta=0$ have been made. This means
that the first quasistatic information state approximately appears
within the first microsecond of the environment deviating from
equilibrium, and so on, as they are marked on the abscissa of
Fig.\ 2. The physical solutions in Eq.\ (\ref{xnt}) with the above
numerical substitutions, take the (positive only) values
$x_n(1.33\times10^{-28}{\rm J}) = \tau({\rm\mu s}) -
1.58\sqrt{n}$, where the eigen-information $x_n$ is given in units
of $1.33\times10^{-28}$\,J when $\tau$ is measured in
microseconds. Furthermore, the minimum information well-depth
required to support the $n^{\rm th}$ state is obtained by
substituting $\tau_n$ in $|V(\tau)|$, i.e.\
$|V(\tau_n)|=\sqrt{(n+\delta)\pi\hbar V_0/t_0}\approx
\sqrt{n}\,(2.1\times 10^{-28}$\,J), which are marked on the
ordinate of Fig.\ 2. In Fig.\ 2 the dynamics of eigen-informations
is plotted for the exactly solvable linear information well.
Clearly, the first information state is formed at $\tau_1$, the
second at $\tau_2$, and so on. This is indeed reminiscent of
particles (as physical realizations of information states) one by
one falling out of equilibrium from the primordial radiation in
the early universe whose phenomenology was discussed in the
introduction. Although qualitatively consistent with the cosmic
evolution phenomenology in the early universe, we must caution
that the numerical parameters used here are not those of the early
universe in the radiation era. It must also be emphasized that any
one type of identical particles in the universe must correspond to
just one information eigen-value. For example all electrons being
identical particles must have the same information content and
therefore correspond to the same eigen-information. The
information states depicted in Fig.\ 2 are examples of
quasistatically evolving or simply {\em quasistatic information
states}, each with a rate of change of the information content
$\dot{x}_n(\tau)$ that sets a {\em speed limit} on the rates of
change of the properties of any physical realization associated
with it \cite{time-info}. It is a matter of considerable interest
that the linear information well with $\alpha = 1$ considered here
sets a threshold value at which this speed limit
$\dot{x}_n(\tau)=V_0/t_0$ is a constant independent of $\tau$. For
$\alpha < 1$ this speed limit is expected to decrease while for
$\alpha > 1$ increase with the time $\tau$. This means that for
the particular case of $\alpha = \frac{4}{3}$ in the matter
dominated era, the evolutionary changes to the information states
are expected to speed up as time passes.

It must be emphasized that the order of the eigen-informations is
such that at any epoch $n=1$ identifies the deepest information
state with the highest eigen-value $x_1(\tau)$. It is the most
complex state of its time with the highest information content. On
the other hand, the smallest eigen-information corresponds to
$n_{\rm max}$, which identifies the most primitive of the
elementary particles with the lowest information content.

\begin{figure}[H]
\centering
\includegraphics[width =0.45\textwidth]{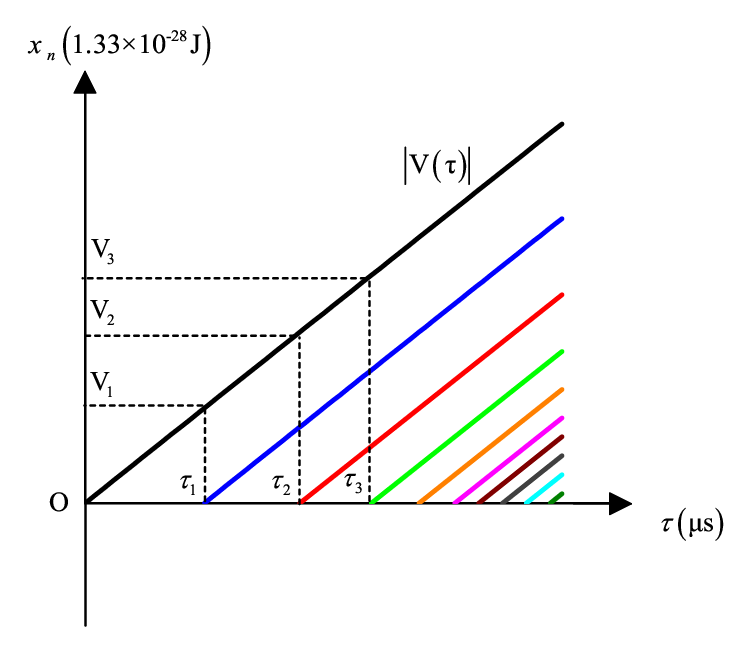}
\caption{The eigen-informations $x_n(\tau)\equiv |I_n(\tau)|$ as a
function of time in the early stages of the deviation from
equilibrium for the exactly solvable linear information well are
shown to illustrate qualitatively the emergence and gradual
evolution of eigen-informations in early universe. Clearly new
information states (shown in blue, red, green, etc.) begin to
emerge one by one at $\tau_n\approx \sqrt{n}$\,($1.58\,\mu$s) when
the information well-depth (shown in black) reaches
$|V(\tau_n)|\approx \sqrt{n}\,(2.1\times 10^{-28}$\,J).}
\label{Fig2}
\end{figure}

\subsection{Present epoch}
In the matter dominated era the power-law exponent for the
information well of Eq.\ (\ref{alpha}) is $\alpha = \frac{4}{3}$
as shown in Fig.\ 1. In cosmological applications at the present
epoch it is often convenient to work with the red-shift parameter
$z$ instead of the time $t$. Using $1+z=a^{-1}(t)$,
$dz=-(\dot{a}/a^2)\,dt$, and the Hubble parameter
$H(z)=\dot{a}/a$, for the change of variables from $t$ to $z$ one
gets $dt= -dz/H(z)\,(1+z)$. Eq.\ (\ref{IT}) with $z$ as
integration variable therefore becomes \be\label{z}
\int_{z_n}^{0}\left[I_n - V(z)\right]
\left(\frac{-dz}{H(z)(1+z)}\right) = \frac{1}{2}(n+\delta)\pi\hbar
\ee where, the upper limit $z=0$ corresponds to the present epoch,
the lower limit $z_n$ is given by $I_n = V(z_n)$, and the dynamics
of the universe is contained in Hubble parameter $H(z)$. From
Friedmann equation $H(z) =
H_0[\Omega_{m,0}(1+z)^3+\Omega_{\Lambda,0}]^{1/2}$, where
$\Omega_{r,0}\simeq 0$ and $\Omega_{k,0}\simeq 0$ have been
neglected. With $\Omega_{m,0}+\Omega_{\Lambda,0}=1$ and
considering {\em matter-only} Friedmann model with
$\Omega_{\Lambda,0}=0$, the Hubble parameter finally becomes
$H(z)=H_0 (1+z)^{3/2}$, where $H_0 = 70\,{\rm km/s/Mpc}=2.27\times
10^{-18}$\,$s^{-1}$ is the Hubble constant. On substituting for
$H(z)$ and $V(z) = -V_0 a^{-1}/(H/H_0)^2 =-V_0 /(1+z)^2$ in Eq.\
(\ref{z}), the following algebraic equation for the information
eigen-values $x_n\equiv -I_n=|I_n|$ at the present epoch is
obtained:
\be\label{xn}\frac{8}{21}\left(\frac{x_n}{V_0}\right)^{7/4}-\frac{2}{3}\left(\frac{x_n}{V_0}\right)+\frac{2}{7}-
\frac{\pi}{2}\frac{\hbar H_0}{V_0}(n +\delta)=0.\ \ee It is worth
noting that essentially the same result can be obtained from Eq.\
(\ref{algebra}) with $\alpha = \frac{4}{3}$, $\tau = t_0$, and
$H_0\sim t_0^{-1}$. Eq.\ (\ref{xn}) gives a {\em static} picture
of the quasistatic information states at the present epoch as
shown compactly in Fig.\ 3. The total number of quasistatic
information states supported by the universe at this epoch is
obtained from Eq.\ (\ref{xn}) with $x_n=0$, which gives $n_{\rm
max}=4.36\times10^{46}$ for the present epoch. Although very
large, the total number of eigen-informations supported by the
universe at the present epoch is {\em finite and countable}. It
must be noted that the ratio $V_0/\hbar H_0 = 2.4\times10^{47}$ in
Eq.\ (\ref{xn}) represents the information well-depth $V_0$ in
units of the smallest quantum of energy at the present epoch
$\hbar H_0 = 2.4\times10^{-52}$\,J \cite{Alf}. As expected, and
much unlike the early stages shown in Fig.\ 2, the spectrum of
eigen-informations at the present epoch is {\em quasi-continuous}.

Figure 3 shows the normalized information potential well
$V(z)/V_0$ at the present epoch with some $I_n$'s shown at
representative intervals $n=10^{46}, 10^{45}, 10^{44}, \cdots$
from top to bottom. The inset shows a semi-log plot of the
normalized quasi-continuous eigen-informations $x_n/V_0$, which at
about $n=10^{45}$ makes a sharp turn towards a plateau that
converges to unity. This is interpreted as a turn of events
towards a radically more complex era within the matter dominated
era that occurs at a time given by
$V(t_{10^{45}})=I_{10^{45}}=-0.84V_0$, which solves to give
$t_{10^{45}}=(0.84)^{\frac{3}{4}} t_0=0.88 t_0$. It is a matter of
considerable interest to note that this time corresponds to 1.7
billion years ago when for the first time multi-cellular life
forms appeared on the earth and the {\em life era} or biological
evolution began in its earnest. Evidently, evolution can be more
predictable than previously thought \cite{Don}, and that
biological evolution can be regarded as a special case of a more
general cosmic evolutionary scenario \cite{Chaisson}. Furthermore,
the transition from non-living to living species as depicted in
the inset of Fig.\ 3 is best characterized by a gradual cross-over
rather than an abrupt phase transition, which means that there is
no life force separating the living from the non-living but their
difference is primarily due to their vastly different information
contents or eigen-informations. Finally, it must be noted that the
exponent $\alpha=\frac{4}{3}$ ($>1$) in the matter dominated era
indicates that, unlike the case of linear information well shown
in Fig.\ 2, the rates of change of the quasistatic
eigen-informations must increase with the time in matter dominated
era and should be at its highest ever at the present epoch, which
is consistent with the most recent studies finding accelerated
changes in all studied animal genomes at the present time
\cite{Bon}.

\begin{figure}[H]
\centering
\includegraphics[width =0.45\textwidth]{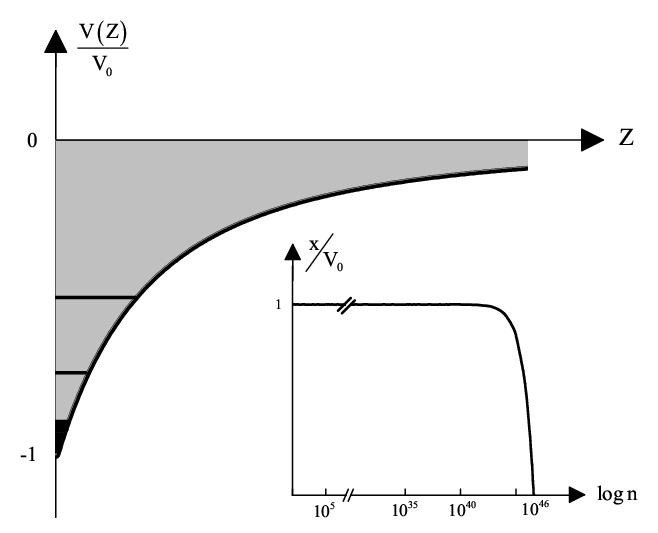}
\caption{The normalized information well at the present epoch
$V(z)/V_0 = -(1+z)^{-2}$ as a function of red-shift parameter $z$.
The spectrum of eigen-informations is quasi-continuous with some
representative $I_n$'s drawn explicitly as horizontal lines for
$n=10^{46}, 10^{45}, 10^{44}, \cdots$ from top to bottom. Inset
shows the semi-log plot of the normalized eigen-informations
$x_n/V_0$ as a function of $\log n$ with a characteristic sharp
turn to a plateau of complexity at about $n=10^{45}$, which
corresponds to the commencement of the life era.} \label{Fig3}
\end{figure}

\section{Summary and final remarks}\label{sec4}
To summarize, therefore, an all new information-time quantization
rule for quasistatic information states in nonequilibrium
thermodynamics based on (i) Shannon information in nonequilibrium
systems, (ii) semiclassical energy-time quantization rule, and
(iii) quasistatic information-energy correspondence, is presented
that must be generally applicable to any slowly varying
nonequilibrium environment where the distance from equilibrium and
the mean temperature are known as a function of time. This all new
quasistatic information-time quantization rule is applied to
cosmic evolution in the matter dominated era (albeit at a
mean-field level) by employing a global thermodynamic potential
for information with phenomenologically determined parameters. It
is  a matter of considerable interest that the results obtained
from this procedure are in agreement with the general
phenomenology of the cosmic evolution at a quantitative level. In
particular a turn of events to the life era or biological
evolution is detected by the theory starting about $1.7$ billion
years ago that is in fact when the multi-cellular life forms
appeared on the earth of which we are only one of the descendants.
As a result biological evolution must be regarded as a special
case of a more general cosmic evolutionary scenario that
encompasses both animate and inanimate species as material
realizations of the quasistatic information states in
nonequilibrium environments \cite{Chaisson}. In our theory,
therefore, the persistent structures (it) are just the material
realizations of the quasistatic information states (bit) supported
by the universe, which is best epitomized by the late Wheeler's
famous catch phrase ``{\it it
from bit}", or to use his precise words \cite{Wheeler}\\
``{\it Otherwise stated, every physical quantity, every it,
derives its ultimate significance from bits, binary yes-or-no
indications, a conclusion which we epitomize in the phrase, it
from bit.}"

Furthermore, the semiclassical information-time quantization rule
introduced here, opens new avenues for further research. Indeed a
more comprehensive application to cosmic evolution must also
involve the inclusion of the radiation dominated era of the early
universe, which can be included in the calculations through its
appropriate thermodynamic potential for information. This must
lead to a quantitatively accurate time line for the birth of
elementary particles, nuclei, light atoms, etc., a qualitative
description of which can be already seen in Fig.\ 2. Such a
calculation may also detect other elementary information states
that are candidates for dark matter, in addition to the better
known baryonic matter. Another significant direction for further
research arises from the fact that past history plays a
significant role in future evolution through the thermodynamic
potential for information in the quasistatic information-time
quantization rule. As a result of this thermodynamic memory, long
after the environment reaches the state of thermodynamic
equilibrium the quantized information states can continue to exist
within the environment, which is to say that quantum processes can
continue for a long time after equilibration due to the past
deviations from equilibrium registered in the memory. This also is
consistent with the recently proposed so-called second-law of
quantum complexity \cite{Susskind}, which asserts that quantum
processes do persist in a system for a long time after
equilibration until they too die out. Finally it must be noted
that the quasistatic information-time quantization rule presented
here constitutes the first step towards a quantum mechanical
theory of information in nonequilibrium thermodynamics, which may
entail time-evolution operators acting on information states,
information operator and information-time uncertainty relation,
etc.


\acknowledgements{One of the authors (SD) is indebted to Prof.\
Brian D.\ Josephson for a number of illuminating discussions and
communications during his visit to the University of Cambridge,
Theory of Condensed Matter (TCM).}


\end{document}